\documentclass[twocolumn]{aastex631}
\shortauthors{Calamari et al.}
\usepackage{amsmath}
\usepackage{mathrsfs}
\usepackage{natbib}
\usepackage{xcolor}

\begin{document}

\title{Predicting Cloud Conditions in Substellar Mass Objects Using Ultracool Dwarf Companions}
%The Utility of Benchmark Companion Brown Dwarfs in Assessing Oxygen Sequestration and Condensate Predictions
\author[0000-0002-2682-0790]{Emily Calamari}
\affiliation{The Graduate Center, City University of New York, New York, NY 10016, USA}
\affiliation{Department of Astrophysics, American Museum of Natural History, New York, NY 10024, USA}

\author[0000-0001-6251-0573]{Jacqueline K. Faherty}
\affiliation{Department of Astrophysics, American Museum of Natural History, New York, NY 10024, USA}

\author[0000-0001-6627-6067]{Channon Visscher}
\affiliation{Department of Chemistry and Planetary Sciences, Dordt University, Sioux Center, IA, USA}
\affiliation{Center for Exoplanetary Systems, Space Science Institute, Boulder, CO, USA}

\author[0000-0002-8871-773X]{Marina E. Gemma}
\affiliation{Department of Geosciences, Stony Brook University, Stony Brook, NY 11794, USA}
\affiliation{Department of Earth and Planetary Sciences, American Museum of Natural History, New York, NY 10024, USA}

\author[0000-0003-4600-5627]{Ben Burningham}
\affiliation{Centre for Astrophysics Research, Department of Physics, Astronomy and Mathematics, University of Hertfordshire, Hatfield AL10 9AB, UK}

\author[0000-0003-4083-9962]{Austin Rothermich}
\affiliation{The Graduate Center, City University of New York, New York, NY 10016, USA}
\affiliation{Department of Astrophysics, American Museum of Natural History, New York, NY 10024, USA}

\begin{abstract}
We present results from conducting a theoretical chemical analysis of a sample of benchmark companion brown dwarfs whose primary star is of type F, G or K. We summarize the entire known sample of these types of companion systems, termed “compositional benchmarks", that are present in the literature or recently published as key systems of study in order to best understand brown dwarf chemistry and condensate formation. Via mass balance and stoichiometric calculations, we predict a median brown dwarf atmospheric oxygen sink of $17.8^{+1.7}_{-2.3}\%$ by utilizing published stellar abundances in the local solar neighborhood. Additionally, we predict a silicate condensation sequence such that atmospheres with bulk Mg/Si $\lesssim$ 0.9 will form  enstatite (MgSiO$_3$) and quartz (SiO$_2$) clouds and atmospheres with bulk Mg/Si $\gtrsim$ 0.9 will form enstatite and forsterite (Mg$_2$SiO$_4$) clouds. Implications of these results on C/O ratio trends in substellar mass objects and utility of these predictions in future modelling work are discussed.

\end{abstract}

\section{Introduction} \label{sec:intro}
Since the first brown dwarf was confirmed in 1995 \citep{Rebolo1995, Nakajima1995}, astronomers have worked to establish a set of spectral standards \citep{Kirkpatrick2005} -- objects whose features define a given spectral type -- as well as characteristic standards \citep[e.g.][]{Pinfield2006, Faherty2010, Gagne2015} -- objects for which properties like age and/or metallicity are independently known. These standards outline and define the changing temperature structures and chemical processes that occur in ultracool dwarf (UCD) atmospheres. Defining this sequence of standards is essential in order to calibrate and test cutting edge evolutionary, photometric and spectral models \citep[ex:][]{Burningham2009, Burningham2013}. While our understanding of brown dwarf science has advanced significantly in the near 30 years since the first spectral standards were discovered, countless open questions remain about the physical and chemical nature of these objects. To name a few: what types of convective processes and/or thermochemistry drives differences between observed and model spectra? What causes the appearance of observable atmospheric clouds and/or variability in certain objects? How can we determine the formation history of an object from observational data? While this work does not attempt to put forth definitive answers to these questions, we do propose a path forward for those who seek to do so through careful examination of brown dwarf characteristic standards, known as “benchmarks".

Individual characterization of brown dwarf atmospheres and fundamental parameters (age, mass, effective temperature ($T_{\rm eff}$), radius) is impeded by the age-mass-temperature degeneracy that exists due to the thermal evolution of these objects. Benchmark brown dwarfs provide the potential to break this degeneracy with the use of independently known parameters. One particularly useful subset of benchmarks are co-movers and companions -- those that (a) belong to a moving group or cluster where the bulk properties of the collection of stars (e.g. age, metallicity, dynamic history) can be inferred \citep[e.g.][]{Gagne2015, Faherty16, Bowler2015b} or (b) are co-moving in binary or multiple star systems where a detailed study of the primary companion star(s) can provide information for each object in the system \citep[e.g.][]{Pinfield2006, Faherty2010, Crepp2012, Deacon2014}. Our focus in this work is on benchmark brown dwarfs belonging to a binary or multiple system with the specification that the primary star in the system is of spectral type F, G or K. In doing so, we aim to map the chemistry of nearby UCDs to the solar neighborhood population ($\leq 100 pc$).

Chemical mapping is one route in exploring the possible origins and evolutionary pathways of stars and planets within our galaxy. This approach has been taken by galactic cartographers to understand large scale chemical structure of disk stars \citep[e.g.][]{Twarog1980, Hawkins2023} as well as solar neighborhood cartographers focusing on exoplanet host star abundances \citep[e.g.][]{Adibekyan2012, Teske2019, Delgado2021}. The chemical characterization of local solar-type (FGK) stars has several benefits -- this data is grounded against the precise chemical abundances of our sun \citep[i.e.][]{Lodders2021}, these spectra are free from molecular absorption bands that make M-type dwarfs difficult to characterize \citep[see][]{Tsuji2014, Tsuji2015} and meteoritic data has established a near 1:1 relationship between solar abundances and primitive condensed material in the protoplanetary disk \citep{Lodders2003, Lodders2021}. 

To ground what types of chemistry we might expect in UCD atmospheres, \citet{Brewer2016b} found that local ($\sim$ $<$ 350 pc) FGK stars have a carbon-to-oxygen (C/O) ratio distribution peaked around 0.47 (lower than that reported for the Sun at 0.55 by \citet{Lodders2021}) with a steep drop off at supersolar values such that no stars in their sample had a C/O $>$ 0.7. This leaves the potential for carbon-rich atmospheres in the solar neighborhood (C/O $>$ 0.65) to be $<$ 0.13$\%$. Additionally, \citet{Brewer2016b} found that the magnesium-to-silicon (Mg/Si) ratio peaks around 1.0 with a broad distribution ranging from 0.8-1.4, where roughly 60$\%$ of systems have 1 $<$ Mg/Si $<$ 2. To understand the population of nearby brown dwarfs it is extremely useful to place them in context with what we currently know about the chemical distribution of the solar neighborhood, and how this agreement or disagreement can inform UCD atmospheric chemistry and formation pathways.

With the use of the spectral inversion, or “retrieval", modeling technique \citep{Line2015, Burningham2017}, we have been able to probe the atmospheres of brown dwarfs and approximate molecular abundances from spectral data. However, even with these methods, converting from atmospheric molecular to bulk atomic abundance is a nontrivial task as we know metal condensates (MgSiO$_3$, Mg$_2$SiO$_4$, Al$_2$O$_3$, etc.) will sequester atoms into clouds starting from the high photosphere down into the deep, unobservable interior. A major participant of this thermochemical process is oxygen, due to its relatively high abundance in these atmospheres and its ability to form both volatile species (e.g. CO, CH$_4$, H$_2$O, etc.) and refractory condensates (e.g. Mg$_2$SiO$_4$, MgSiO$_3$, Al$_2$O$_3$, etc.). As a result, retrieving a reliable oxygen abundance for UCD atmospheres has been challenging \citep[see e.g.,][]{Line2017, Calamari2022, Zalesky2022}. To date, the attempts to account for oxygen sequestered in unobservable condensates have relied on an approximate correction factor from solar-ratio, thermochemical calculations done in \citet{burrowssharp1999} (in which the dominating oxygen sink for atmospheres $<$ 2200 K is enstatite (MgSiO$_3$) clouds), or the estimated removal of $\sim$ 20-23\% of bulk oxygen based upon the abundances of rock-forming elements in a solar-composition gas \citep{Visscher2005,visscher2010icarus}. The inability to accurately measure oxygen abundances directly impacts all of the aforementioned open questions in brown dwarf science; inhibiting not only understanding of the thermochemical and dynamical processes that govern these atmospheres, but also measurements of fundamental properties like metallicity and C/O ratio that may reveal formation history.

To focus on answering some of these open questions, we revisit the work of earlier chemical models \citep{Allard1997, burrowssharp1999, Ackerman2001a, Marley2002, Visscher2005, Visscher2010, visscher2010icarus} and update these assumptions to include an oxygen sink correction that is reflective of non-solar abundances and considers other minor refractory condensates that could shape observable spectra. By focusing on brown dwarf systems co-moving with a solar-type star and utilizing the bulk chemistry known for the primaries in these benchmark systems, we estimate condensate species and abundance based on a given system's unique stellar, rather than solar, bulk abundances. For this theoretical framework, we will employ mass balance constraints and stoichiometric procedures found in previous thermochemical equilibrium studies of condensate formation in UCD atmospheres \citep{Lodders2002, Visscher2005, Visscher2010, wakeford2017}. By using a more accurate assessment of the chemical context of stellar systems, we aim to provide a more rigorous guide to estimating the loss of oxygen to clouds in UCD atmospheres by taking advantage of the full data set on benchmark brown dwarfs.

In \autoref{sec:benchmarks}, we discuss the utility of benchmark brown dwarfs in addressing current open questions in brown dwarf science and present the updated summary of companion benchmark systems (\autoref{sec:compositional benchmark}). In \autoref{sec:l dwarfs}, we discuss the importance of L-type dwarf benchmarks in constraining atmospheric chemistry as it relates to cloud model solutions and their impact on UCD modelling as a whole. In \autoref{sec:sample}, we outline the chosen published stellar abundance data set and the subsequent subset of companion benchmarks highlighted. In \autoref{sec:spectra}, we give an overview of the theoretical gaseous absorbers and condensate species we expect to find in UCD atmospheres. In \autoref{sec: framework}, we outline the theoretical thermochemical framework and assumptions used in order to carry out this work. In \autoref{sec: predictions}, we present our results from applying thermochemical equilibrium calculations to a collection of abundances of solar neighborhood FGK-type stars in order to quantify major oxygen sinks in companion UCD atmospheres. In \autoref{sec: future work}, we discuss the implications of these findings on future UCD atmospheric modelling.

\section{Significance of Benchmark Brown Dwarfs}\label{sec:benchmarks}
The first confirmed methane-bearing brown dwarf was Gliese 229B, which was found co-moving with the nearby M dwarf Gliese 229A.  The kinematics and general chemical make-up of Gliese 229A were known at the time, therefore Gliese 229B was marked a ``benchmark" brown dwarf given that the properties of the primary could be applied to the secondary. The term benchmark is now used more generally to highlight a UCD for which we have external empirical constraints that do not rely on model predictions \citep[e.g.][]{Pinfield2006, Burningham2009}. By 2010, approximately 70 benchmarks were discovered and catalogued in the literature \citep[e.g.][]{Burgasser2005b, Faherty2010, Seif2010, Bihain2010} -- a number that has more than doubled in recent years thanks to large or all-sky surveys such as the Two-Micron All-Sky Survey \citep[2MASS;][]{2MASS}, Wide-Field Infrared Survey Explorer \citep[WISE;][]{WISE} and Panoramic Survey Telescope and Rapid Response System \citep[Pan-STARRS;][]{Panstars} as well as citizen science projects like the Backyard Worlds Planet 9 Collaboration \citep[see][]{Kuchner2017, Faherty2021, Schneider2021, Rothermich2023}. This subset of brown dwarfs has systematic or spectral information that can lead to a more precise age estimation, metallicity and/or mass -- all of which are extremely difficult parameters to probe independently for UCDs.

The largest (and fastest growing) sample of benchmark objects comes from those that are widely separated from main sequence stars due to their relative abundance and ease of identification as they are individually resolved from their host star. At present there are $>$ 175 known companion benchmark brown dwarfs catalogued in the literature \citep[see][]{Faherty2010, Deacon2014, Rothermich2023}. To understand the particular benefit of benchmark brown dwarfs, we can look to studies that have shown how empirical constraints can test current atmospheric and evolutionary models. For example, \citet{Burningham2009} presented the discovery of Wolf 940B, a T8.5 companion to the M4 dwarf, Wolf 940. In this work, they were able to take robust measurements for distance, metallicity and age from the primary and apply these to the secondary T dwarf companion to test model repeatability with a set of solar metallicity BT-Settl models \citep{Allard2003}. They found that the models underestimated the flux peak in the K spectral band resulting in an overestimated model temperature of $\sim$ 100 K. They also showed that, at the time, non-solar metallicity model spectra did not agree well with observed data for late T dwarfs. With benchmark systems involving FGK-type primaries, we can even more precisely use inferred chemical properties from the primary (metallicity and abundance) to examine UCD atmospheric and evolutionary models such as work done in \citet{Bowler2009, Mann2013, Line2015, Wang2022}.

These works have been able to point out incongruities, or, alternatively, consistencies, between observed data and spectral and evolutionary models so that we may understand and tweak our underlying theoretical physical assumptions accordingly. While spectral or evolutionary analysis on non-benchmark brown dwarfs can also shed light on model precision, it is much more challenging to pinpoint the cause of poor matches when there are no certain constraints. This remains the significant motivation for studying benchmark brown dwarfs.

\subsection{The Compositional Benchmark Sample}\label{sec:compositional benchmark}
Of the collection of co-moving benchmark brown dwarfs, the main employment of known empirical properties is done through posterior comparison to resulting model values. While this is a useful comparative tool, it can also leave us with more open questions when known and model values are not within statistical agreement. As discussed above, this has been done in \citet{Burningham2009} to comment on and test model reproducibility of known spectral features. More recently, it can be seen in the work of \citet{Line2015, Kitzmann2020, Calamari2022} which all utilized versions of retrieval methods \citep[see][etc.]{Line2015, Burningham2017, Moll2020} to analyze the observed spectra of Gl 570D and HD 3651B \citep{Burgasser06, Mugrauer2006}, Epsilon Indi Bab \citep{King2010} and Gl 229B \citep{Nakajima1995, Oppenheimer1995}, respectively, and determine the highest likelihood fundamental parameters (i.e. molecular abundances, $T_{\rm eff}$, log(g), [M/H], etc.). To varied statistical degrees, what was reported in these studies was an inconsistency between the retrieved metallicity and carbon-to-oxygen (C/O) ratio of the companion and that known for its stellar host.

This application of benchmark data has proven to be a powerful tool in examining and challenging the assumptions we might have about a particular system. Specifically, in \citet{Calamari2022} the retrieved carbon and oxygen abundances in Gl 229B were compared to those reported for its primary, revealing that Gl 229B appeared to be comparatively oxygen depleted ($\approx$ 3$\sigma$ disagreement). While there are a few ways of interpreting this result - most notably that the nascent origins of these two companions are inherently different - \citet{Calamari2022} concluded that this discrepancy, in part, was suggestive of misunderstood cloud chemistry in the modelling and theoretical estimations. It is important to note that only carbon and oxygen abundances for Gl 229A were readily available for comparative use in this system. However, this alone provided insight and motivation to re-examine the ways in which we model brown dwarf atmospheres and account for clouds.

To fully utilize companion benchmarks in our modeling requires systems in which a wider range of atomic abundances are readily available. To predict the condensate species and cloud particle density, we need a picture of the chemical landscape beyond just carbon and oxygen. Specifically, we want to know the abundances of the dominant reactive metal species (Mg, Si, Al, and Ca) to determine what kinds of clouds we would expect to see if the UCD companion did, in fact, contain the same elemental abundances as its stellar companion. While most co-moving brown dwarfs are found orbiting M-dwarf stars \citep{Faherty2010, Deacon2014}, it is notoriously difficult to calculate atomic abundances for stars of this spectral type due to substantial molecular absorption bands throughout their spectra. As a result, we turn to benchmark systems in which the primary star is of spectral type F, G or K in hopes of attaining such chemical information.

While there have been several individual discoveries and compiled samples of benchmark brown dwarfs \citep[e.g.][]{Pinfield2006, Burningham2009, Bowler2009, Faherty2010, Deacon2014, Rothermich2023}, here we conduct a thorough literature search for all brown dwarfs co-moving with an F, G or K primary. While this criterion significantly limits the sample size, as more than half of all known co-moving brown dwarfs are companions to objects of spectral type M or later, we do this to prioritize the availability and accuracy of stellar atomic abundances. In \autoref{tab:compositional benchmarks}, we list all known and newly discovered brown dwarfs co-moving with an F, G or K primary star, labeling this subset as the brown dwarf compositional benchmark sample. It is important to note that in classifying compositional benchmarks, attention has been taken to isolate secondaries that are members of unresolved binaries, as disentangling their combined light spectra is a nontrivial task and can complicate analysis. One example of this type of system would be Gl 417 BC which is a spectroscopically unresolved L4.5 + L6 brown dwarf binary comoving with the G2 dwarf, Gl 417, \citep{Kirkpatrick2001, Dupuy2014}. However, as there is an increased binary frequency among widely separated brown dwarfs to stellar-type primaries \citep{Burgasser2005}, we include these here for completeness. 
%as well as eclipsing binary systems like the L6 + M5 system, GJ 802 AB \citep{Knapp2004, Pravdo2005}.

\begin{longrotatetable}
\begin{deluxetable*}{ccccccccccr}
\rotate
\tabletypesize{\scriptsize}
\tablenum{1}
\tablecaption{The Compositional Benchmark Sample\label{tab:compositional benchmarks}}
\tablehead{\colhead{Object} & \colhead{Primary} & \colhead{RA} & \colhead{Dec} & \colhead{SpT} & \colhead{SpT} & \colhead{$d_{Primary}$} & \colhead{Separation} & \colhead{Proj. Sep} & \colhead{Age} & \colhead{Ref}\\
\colhead{} & \colhead{} & \colhead{} & \colhead{} & \colhead{Secondary} & \colhead{Primary} & \colhead{(pc)} &  \colhead{(arcsec)} & \colhead{(AU)} & \colhead{(Gyr)} & \colhead{}}
\startdata
    \textbf{Known Systems} &  &  &  &  &  &  &  &  & & \\
    \hline
    2MASS J00193275+4018576 & LP 192-58 & 4.8859931 & 40.3151441 & L2 & K7 & 55.21 $\pm$ 0.11 & 58.5 & 3990 & 0.3-10 & 1, 2 \\
    2MASS J00302476+2244492 & BD+21 55 & 7.6042199 & 22.7461537 & L0.5 & K2 & 37.91 $\pm$ 0.14 & 117.1 & 3970 & 0.5-10 & 1, 2 \\
    HD 3651 B & HD 3651 & 9.829614 & 21.254559 & T7.5 & K0.5V & 11.14 $\pm$ 0.01 & 43 & 480 & 0.7-4.7 & 1, 2, 8, 9, 23 \\
    HD 4113 C & HD 4113 & 10.8025 & -37.9826306 & T9 & G5V+M1V & 41.92 $\pm$ 0.09 & 0.535 & 22 & 3-6 & 27 \\
    HD 4747 B & HD 4747 & 12.361505 & -23.212463 & L/T & G8/K0V & 18.85 $\pm$ 0.01 & 0.61 & 10 & 0.9-3.7 & 18 \\
    ULAS J014016.91+015054.7 & BD+01 299 & 25.071311 & 1.8484382 & T5 & K5 & 38.56 $\pm$ 0.03 & 31 & 35-45 & 6.5-13.5 & 3, 12 \\
    2MASS J01591078+3312313 & HD 12051 & 29.7959843 & 33.207192 & L6 & G9V & 24.77 $\pm$ 0.02 & 52.1 & 1300 & 2.2-10.2 & 1, 2 \\
    HD 13724 B & HD 13724 & 33.086156 & -46.816377 & T4 & G3/5V & 43.48 $\pm$ 0.06 & 0.24 & 26.3 & 0.05-1.5 & 26 \\
    2MASS J02233667+5240066 & HD 14647 & 35.902796 & 52.668514 & L1.5 & F5 & 80.41 $\pm$ 0.46 & 47.7 & 3300 & 0.5-2.4 & 2 \\
    2MASS J02355993-2331205 & HD 16270 & 39.00019195 & -23.52224277 & L1 & K2.5Vk & 21.22 $\pm$ 0.02 & 11.95 & 250 & <1 & 1, 3 \\
    HD 19467 B & HD 19467 & 46.827 & -13.762028 & T5.5 & G3V & 32.03 $\pm$ 0.02 & 1.6 & 51 & 4-10 & 13 \\
    HIP 21152 B & HIP 21152 & 68.019917 & 5.409944 & L/T & F5V & 43.21 $\pm$ 0.05 & 408 & 17.5 & 0.65-0.85 & 1, 29\\
    HD 33632 Ab & HD 33632 Aa & 78.3208	& 37.2808 & L9.5 & F8V & 26.39 $\pm$ 0.02 & 0.75 & 20 & 1.2-4.5 & 19 \\
    2MASS J05394952+525352 & HD 37216 & 84.9566683 & 52.8992533 & L5 & G5V & 28.08 $\pm$ 0.04 & 27 & 753 & 1.1-9.3 & 1, 2 \\
    2MASS J06135342+1514062 & HD 253662 & 93.4725944 & 15.234332 & L0.5 & G8IV & 86.46 $\pm$ 0.34 & 20.1 & $>$ 1252 & $<$ 10 & 1, 2 \\
    AB Pic B & AB Pic & 94.804162 & -58.055611 & L1 & K1V & 50.14 $\pm$ 0.03 & 5.5 & 250-270 & 0.03 & 1, 16, 23 \\
    2MASS J06324849+5053351 & LSPM J0632+5053 & 98.202075 & 50.893106 & L1.5 & G2 & 82.58 $\pm$ 0.13 & 47.4 & 4499 & 0.2-10 & 2 \\
    2MASS J06462756+7935045 & HD 46588 & 101.6121946 & 79.5818179 & L9 & F7V & 18.21 $\pm$ 0.04 & 79.2 & 1420 & 1.3-4.3 & 1, 2 \\
    HD 47197 B & HD 47197 & 102.339167 & 43.759194 & L4 & F5V & 41.47 $\pm$ 0.05 & 0.8 & 43 & 0.26-0.79 & 1, 15 \\
    2MASS J07580132-2538587 & HD 65486 & 119.5073205 & -25.6508698 & T4.5 & K4Vk & 18.48 $\pm$ 0.01 & 88 & 1630 & 0.3-2.8 & 1, 2 \\
    eta Cnc B & eta Cnc & 128.132502 & 20.449967 & L3.5 & K3III & 97.48 $\pm$ 0.83 & 2.2-3.5 & 154 & 15000 & 1, 17 \\
    HD 72946 B & HD 72946 & 128.963611 & 6.62277 & L5 & G8V & 25.87 $\pm$ 0.08 & 6.5 & 10 & 1-2 & 1, 28 \\
    2MASS J10221489+4114266 & HD 89744 & 155.5623583 & 41.2457764 & L0 & F7V  & 38.68 $\pm$ 0.11 & 2460 & 63 & 1.5-3 & 1, 2, 23 \\
    2MASS J11102921-2925186 & CD-28 8692 & 167.621714 & -29.4221669 & L2 & K5V & 39.79 $\pm$ 0.07 & 50.8 & 2026 & 9.5-13.5 & 1, 25 \\
    2MASS J12173646+1427119 & HD 106888 & 184.4015804 & 14.4531479 & L1 & F8 & 67.18 $\pm$ 0.57 & 38.1 & 2170 & 0.3-2.5 & 1, 4 \\
    WISE J124332.17+600126.6 & BD+60 1417 & 190.88386 & 60.023957 & L8$\gamma$ & K0 & 44.96 $\pm$ 0.03 & 37 & 1662 & 0.01-0.15 & 11 \\ 
    2MASS J13005061+4214473 & BD+42 2363 & 195.2084201 & 42.246548 & L1 & K6V & 44.15 $\pm$ 0.06 & 132.8 & 5640 & 0.3-10 & 1, 2 \\
    GJ 499 C & GJ 499 AB & 196.420872 & 20.7779818 & L4 & K5+M4 & 19.65 $\pm$ 0.02 & 516 & 9708 & 3-5 & 1, 5 \\
    2MASS J13204427+0409045 & HD 116012 & 200.1820776 & 4.1522243 & L5 & K0V & 30.31 $\pm$ 0.046 & 516 & 9708 & 12-14 & 1, 5 \\
    ULAS J13300249+0914321 & TYC 892-36-1 & 202.5102524 & 9.2422718 & L2 & K-type & 246.85 $\pm$ 2.80 & 260.4 & ... & 0.2-1.5 & 1, 4 \\
    2MASS J13324530+7459441 & BD+75 510 & 203.188635 & 74.995628 & L2 & K8 & 35.40 $\pm$ 0.01 & 38.3 & 1364 & 0.2-1.4 & 5 \\
    HD 118865 B & HD 118865 & 204.9323213 & 1.0766982 & T5 & F7V & 60.80 $\pm$ 0.20 & 148 & 9200 & 1.5-4.9 & 1, 2 \\
    2MASS J14165987+5006258 & HD 125141 & 214.2474598 & 50.1080132 & L4 & G5 & 47.11 $\pm$ 0.06 & 570 & ... & 8.5-11 & 1, 34 \\
    ULAS J142320.79+011638.2 & HD 126053 & 215.8371027 & 1.276492 & T8 & G1.5V & 17.44 $\pm$ 0.01 & 152.8 & 2630 & 2.3-14.4 & 1, 33 \\
    2MASS J14284235-4628393 & CD-45 9206 & 217.1761684 & -46.4784943 & T4.5 & K7Vk & 24.07 $\pm$ 0.02 & 377.3 & 9000 & 1-5 & 1, 7 \\
    HD 130948 BC & HD 130948 & 222.566667 & 23.911611 & L4+L4 & G2V & 18.20 $\pm$ 0.01 & 2.64 & 46.5 & 0.4-0.9 & 20, 21 \\
    GJ 570 D & GJ 570 & 224.3175381 & -21.3712191 & T7 & K4V & 5.88 $\pm$ 0.002 & 261.7 & 1525 & 2-10 & 1, 10, 23 \\
    ULAS J150457.65+053800.8 & BD+06 2986 & 226.2388579 & 5.632459 & T6 & K8V & 19.02 $\pm$ 0.02 & 63.8 & 1230 & $>$ 1.6 & 1, 2 \\
    2MASS J15232263+3014562 & * eta CrB & 230.8449168 & 30.2481943 & L8 & G2V+G2V & 17.86 $\pm$ 0.25 & 195.3 & 3635 & 3-5 & 6 \\
    2MASS J17262235-0502110 & * 47 Oph & 261.59315 & -5.0364 & L5.5 & F3V* & 32.27 $\pm$ 0.16 & 294.1 & 1890 & 1.6-1.9 & 2 \\
    2MASS J18005854+1505198 & HD 164507 & 270.2436883 & 15.08842874 & L1 & G5IV & 45.44 $\pm$ 0.07 & 25.5 & 1136 & 3-4 & 1, 25 \\
    %2MASS J19073307+3015304 & V* V478 Lyr & 286.8884528 & 30.258925 & L1 & G6V & 27.12 $\pm$ 0.02 & 17.0 & ... & 11.5-13.5 & 1 \\ 
    HR 7672 B & HR 7672 & 301.025833 & 17.070278 & L4 & G0V & 17.77 $\pm$ 0.01 & 0.79 & 14 & 1-3 & 1, 22 \\
    HD 203030 B & HD 203030 & 319.74572 & 26.22948 & L7.5 & K0V & 39.29 $\pm$ 0.09 & 11 & 487 & 0.13-0.4 & 1, 2, 6, 23 \\
    2MASS J21442847+1446077 & V* HN Peg & 326.1198745 & 14.7683382 & T2.5 & G0V$+$ & 18.13 $\pm$ 0.02 & 42.9 & 795 & 0.1-0.5 & 1, 8, 23 \\
    $\epsilon$ Indi Bab & $\epsilon$ Indi & 331.0767776 & -56.793953 & T1.5+T6 & K5V & 3.64 $\pm$ 0.003 & & 1459 & 0.8-2.0 & 1, 14, 23\\   
    2MASS J22461844+3319304 & BD+32 4510 & 341.576865 & 33.325119 & L1.5 & K2* & 64.68 $\pm$ 3.23 & 16 & 1040 & 0.1-10 & 2 \\
    \hline
    \textbf{Newly Discovered Systems} & &  &  &  &  &  &  &  &  & \\
    \hline
    CatWISE J005635.48-240401.9 & HIP 4417 & 14.1478506 & -24.0672083 & L8 & K0 & 67.60 $\pm$ 0.08 & 102 & 6924 & ... & 1, 24\\
    CatWISE J030005.73-062218.6 & BPS CS 22963-0014 & 45.0238923 & -6.371848 & L9 & K7 & 67.13 $\pm$ 0.08 & 63 & 4200 & ... & 1, 24\\
    CatWISE J055909.00-384219.8 & HD 40781 & 89.787502 & -38.7055027 & L4 & G0V & 60.65 $\pm$ 1.30 & 54.5 & 3259 & $<$ 1 & 1, 24\\
    CatWISE J065752.45+163350.2 & HD 51400 & 104.46857 & 16.563966 & L6 & G5 & 37.08 $\pm$ 0.78 & 64 & 2254 & ... & 1, 24\\
    CatWISE J085131.24-603056.2 & PM J08515-6029 & 132.8801846 & -60.5156128 & L3 & K7 & 30.93 $\pm$ 0.01 & 95.3 & 2948 & $<$ 1 & 1, 24\\
    CatWISE J133427.70-273053.1 & HD 117987 & 203.61543 & -27.514766 & L0 & K3V & 36.95 $\pm$ 0.46 & 50 & 1772 & ... & 1, 24\\
    CatWISE J183207.94-540943.3 & HD 170573 & 278.03311 & -54.162028 & T7 & K4.5Vk & 19.12 $\pm$ 0.01 & 619.3 & 11843 & 9-13.5 & 1, 24\\
    \hline
    \textbf{Known Unresolved Binaries} & &  &  &  &  &  &  &  &  & \\
    \hline
    2MASS J00250365+4759191 & HD 2057 & 6.2669728 & 47.9877566 & L4+L4 & F8 & 54.01 $\pm$ 0.40 & 210 & 8800 & $<$ 1 & 32\\
    HD 8291 B & HD 8291 & 20.5708829 & 3.522572 & L1+T3 & G5V & 50.38 $\pm$ 0.35 & 44.9 & 2570 & 0.5-10 & 1, 2 \\
    Gl 337 CD & Gl 337 & 138.0584919 & 14.9956706 & L8.5+L7.5 & G8V+K1V & 20.35 $\pm$ 0.14 & 43 & 881 & 0.6-3.4 & 1, 31\\
    Gl 417 BC & Gl 417 & 168.1055653 & 35.8028953 & L4.5+L6 & G2 & 22.65 $\pm$ 0.02 & 90 & 2000 & 0.08-0.3 & 1, 30 \\ 
\enddata
\tablerefs{ 1.\citet{Gaia2021}, 2. \citet{Deacon2014}, 3. \citet{Burningham2018}, 4. \citet{Marocco2017}, 5. \citet{Gomes2013}, 6. \citet{Pinfield2006}, 7. \citet{Lodieu2014}, 8. \citet{Luhman2007}, 9. \citet{Mugrauer2006}, 10. \citet{Burgasser2000}, 11. \citet{Faherty2022}. 12. \citet{Skrzypek2016}, 13. \citet{Crepp2012}, 14. \citet{Scholz2003}, 15. \citet{Metchev2004}, 16. \citet{Chauvin2005}, 17. \citet{Zhang2010}, 18. \citet{Crepp2016}, 19. \citet{Currie2020}, 20. \citet{Dupuy2009}, 21. \citet{Potter2002}, 22. \citet{Liu2002}, 23. \citet{Faherty2010}, 24. \citet{Rothermich2023}, 25. \citet{Marocco2020}, 26. \citet{Rickman2020}, 27. \citet{Cheetham2018}, 28. \citet{Maire2020}, 29. \citet{Kuzuhara2022}, 30. \citet{Kirkpatrick2000}, 31. \citet{Wilson2001}, 32. \citet{Reid2006}, 33. \citet{Pinfield2012}, 34. \citet{Chiu2006}}
\end{deluxetable*}
\end{longrotatetable}

\section{The Role of L-Type Dwarfs in Benchmarking}\label{sec:l dwarfs}
At the ultracool end of the spectral sequence, the L, T and Y dwarfs exhibit spectra filled with large molecular absorption bands and atmospheres cool enough to form clouds. Of the 57 systems in the compositional benchmark sample listed in Table ~\ref{tab:compositional benchmarks}, $\sim$ 75$\%$ are systems in which the UCD companion is an L-type dwarf, a spectral classification bounded by 1300 $ < T_{\rm eff} < $ 2200 K. This statistic is likely due to an observational bias as L dwarfs in the local region will be brighter, and therefore easier to identify, than T or Y dwarfs. However, this bias works in our favor in regard to calibrating and tuning atmospheric modelling of brown dwarfs due to the particularly cloudy photospheres of L dwarfs \citep{Burningham2017, Suarez2022, Vos2023}.

\citet{Calamari2022} showed that the population of brown dwarfs modelled with retrievals have an anomalously high median C/O ratio ($\sim$ 0.79) which is shown to be inconsistent with the solar neighborhood, where the median C/O ratio for local F, G and K-type dwarfs is $\sim$ 0.47 \citep{Brewer2016}. While we could consider that this is a real attribute of this sample of brown dwarfs, uniformly oxygen-depleted atmospheres suggests a systematic modelling error in the way clouds are accounted for, as oxygen-rich condensates are known to play a major role in atmospheres $<$ 2200 K. Within the substellar population, the most abundant object choices for evaluating the influence of clouds are L and T dwarfs. T dwarfs are the more simplistic objects to approach given the relative lack of silicate clouds in the photosphere and dominance of methane gas.  However, for the examination of oxygen sequestration and condensate species formation, it is useful to examine the more complex photospheres of L dwarfs.

A notable feature of UCD atmospheres is condensate formation, not only theorized to exist \citep[e.g.][]{Lunine1986, Tsuji1996a, Marley1999a, Ackerman2001a, Lodders2002,Lodders2004science, Kirkpatrick2005} but also evidenced in observable mid-infrared spectra \citep{Cushing2006, Suarez2022}. As described in \citet{Kirkpatrick2005}, early L dwarfs near the M/L transition see the spectral line disappearance of TiO and VO as those molecules are sequestered into oxygen-bearing condensates (e.g., CaTiO$_3$, Ca$_4$Ti$_3$O$_{10}$, Ca$_3$Ti$_2$O$_7$, Ti$_2$O$_3$, Ti$_3$O$_5$, Ti$_4$O$_7$). Related to titanium condensate chemistry are aluminum and calcium, which form condensates (i.e. Al$_2$O$_3$, CaAl$_{12}$O$_{19}$, CaAl$_4$O$_7$, Ca$_2$Al$_2$SiO$_7$) at slightly higher temperatures than titanium, but similarly impact the availability of oxygen \citep{burrowssharp1999, Allard2001, Lodders2002, LodFeg2002, wakeford2017}.

A similar phenomenon occurs with iron, magnesium and silicon at the transition into the mid-L (L4-L6) regime. The FeH and CrH absorption features that shape early- to mid-L spectra begin to weaken with the appearance of iron and magnesium-silicate clouds -- most notably, forsterite (Mg$_2$SiO$_4$) and enstatite (MgSiO$_3$), the Mg-rich endmembers of olivine and pyroxene, respectively \citep{burrowssharp1999, LodFeg2002,loddersfegley2006,lodders2010_cloudchem,Visscher2010}. Mid-infrared observational evidence of this silicate cloud feature is catalogued in \citet{Suarez2022}, which measured the strength of silicate absorption at 8-11 $\mu$m in 69 L dwarfs across spectral type L0-L8. They found evidence for silicate clouds across the L spectral sequence while also noting that this feature disappears upon entering the T spectral type (around T2). Despite theoretical predictions of photospheric alkali metal (KCl, Na$_2$S) and sulfide (MnS, ZnS) clouds in T dwarfs \citep[e.g.,][]{Morley2012}, they generally exhibit cloudless photospheres with the potential for the same silicate clouds to be forming in deeper, unobservable parts of the atmosphere \citep[see][]{Kirkpatrick2005, Line2015, Line2017, Calamari2022}.

Accordingly, if we want to understand and accurately model the chemistry of UCDs, we must first understand the thermochemical processes behind condensate (i.e., cloud) formation. In order to check these thermochemical cloud predictions, a useful starting point is studying L dwarf spectra where we can test theory against observation, making L dwarf chemical benchmarks a prime target of study.

\section{Sample Selection}\label{sec:sample}
In exploring well-characterized systems, we began by cross-matching our compositional benchmark sample with published elemental abundance studies for main sequence stars \citep[e.g.][]{Adibekyan2012, Brewer2016, Delgado2021} in search of a host companion with well characterized chemistry. We prioritized literature spectroscopic chemical abundance studies of main sequence stars that had a uniform observational and/or reduction set-up to minimize systematics which could contaminate the analysis. We found that \citet{Brewer2016} and \citet{Rice2020} were inclusive of our sample and produced robust measurements for temperature, gravity, metallicity and abundance over a large sample ($>$ 2,000) of main sequence stars. Both works utilized high resolution (R$\sim$ 70,000), high signal-to-noise (S/N $\geq$ 200) HIRES spectra from the Keck I telescope. In \citet{Brewer2016}, one-dimensional (1D) local thermodyamic equilibrium (LTE) models were iteratively fit to observed spectra using the procedure described in \citet{Brewer2015}. \citet{Rice2020} added to this work by using \textit{The Cannon}, a machine learning technique \citep{Ness2015, Casey2016}, to build a well-characterized model trained on the data set from \citet{Brewer2016} that was shown to efficiently obtain high-precision stellar parameters with improved speed and accuracy. Both of these studies combined provide a uniform and reliable catalogue of stellar parameters and abundances from which we base our study.

In \autoref{tab:sample}, we outline the compositional benchmarks whose primaries have been thoroughly studied in either \citet{Brewer2016} or \citet{Rice2020}, a total of 12 stars. We use the procedure outlined in \citet{Brewer2016b} to convert from reported $[X/H]$, the ${\rm log}_{10}$ of the solar relative number abundance of an element with respect to hydrogen, to abundance ratios for a given two elements:
\begin{equation}
    X_1/X_2 = 10^{\rm ([X_1/H] - (X_1/H)_{\odot})-([X_2/H] - (X_2/H)_{\odot})}
\end{equation}
We do this for [C/H], [O/H], [Mg/H], [Si/H], [Ca/H] and [Al/H] to determine C/O, Mg/Si and Ca/Al abundance ratios. Additionally, we examine these elemental abundances along with [Ti/H] and [V/H] as required inputs of the thermochemical equlibrium procedures detailed in \autoref{sec: framework}. We use the solar elemental abundances published in \citet{Lodders2021}.

\begin{figure*}[ht!]
\centering
\includegraphics[width=0.8\textwidth]{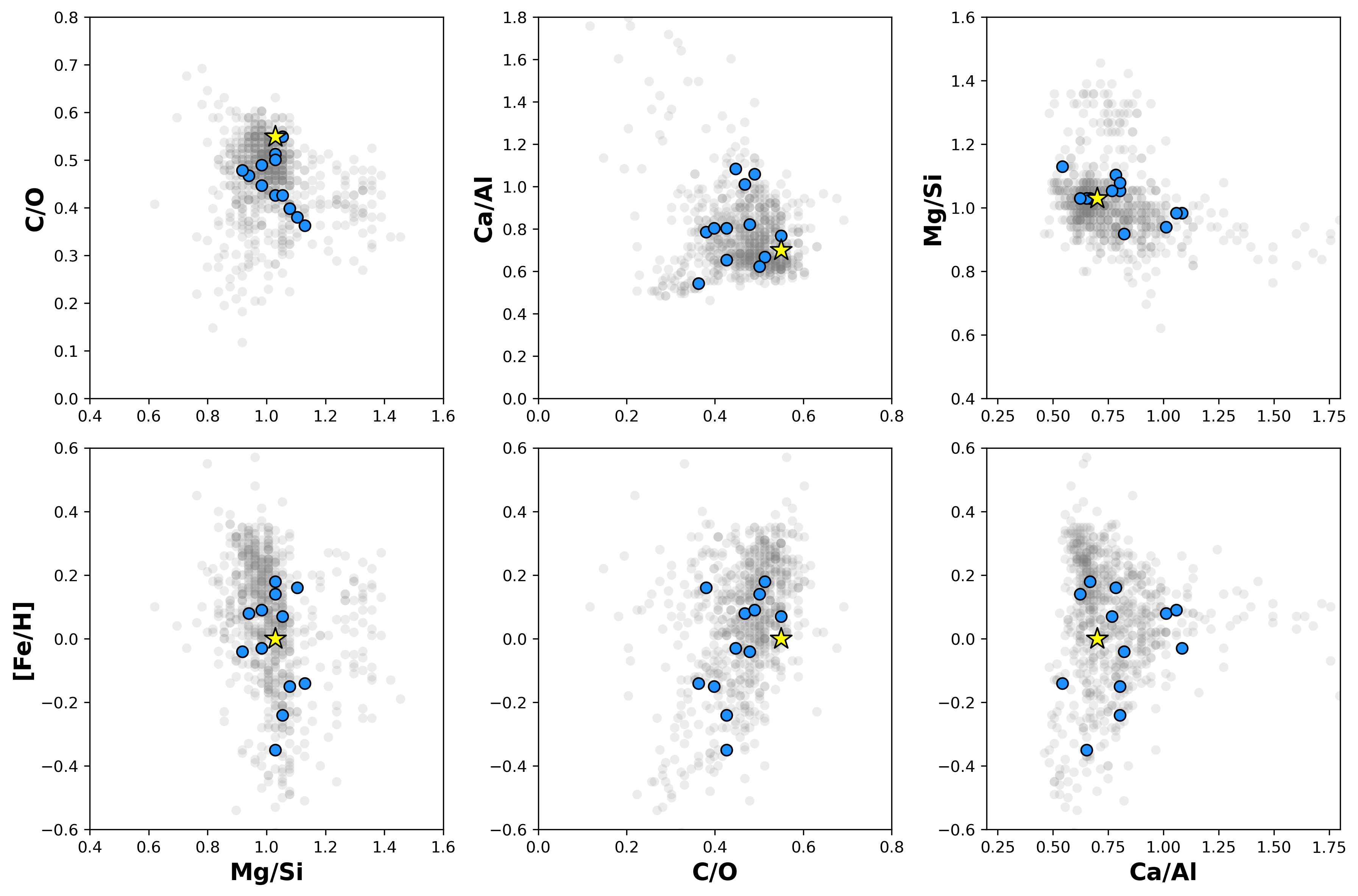}
\caption{Highlighting the abundance ratios of the compositional benchmarks found in the \citet{Brewer2016, Rice2020} stellar abundances catalogues. Blue points represent the compositional benchmarks laid out in \autoref{tab:compositional benchmarks} while the yellow star indicates solar abundance ratios. The grey points show the chemical variance in solar neighborhood FGK stars from \citet{Brewer2016}. \label{fig:sample}}
\end{figure*}

While we highlight this intersection of known compositional benchmarks with elemental abundances from \citet{Brewer2016} and \citet{Rice2020} (12 stars), we use the data set from \citet{Brewer2016} with a cutoff for stars $\leq$ 100 pc (746 stars) as a guide for the kind of chemical distributions we might expect to find across the solar neighborhood. In \autoref{fig:sample}, we show abundance ratios for C/O, Mg/Si and Ca/Al for the compositional benchmark subset and the \citet{Brewer2016} solar neighborhood sample plotted against each other as well as overall metallicity, traced by [Fe/H], to outline the range of local chemical abundances on which we focus our discussion.

\begin{deluxetable}{lcccc}
\tablenum{2}
\tabletypesize{\small}
\tablecaption{Abundance Ratios of Selected Compositional Benchmark Primaries \label{tab:sample}}
\tablehead{\colhead{\textbf{Name}} & \colhead{\textbf{C/O}} & \colhead{\textbf{Mg/Si}} & \colhead{\textbf{Ca/Al}} & \colhead{\textbf{Ref}}}
\startdata
    HD 12051 & 0.513 & 1.030 & 0.668 & 1\\
    HD 203030 & 0.468 & 0.939 & 1.011 & 1\\
    HD 46588 & 0.447 & 0.984 & 1.083 & 1\\
    HD 126053 & 0.426 & 1.030 & 0.653 & 1\\
    HD 19467 & 0.363 & 1.129 & 0.543 & 1\\
    HD 37216 & 0.479 & 0.918 & 0.822 & 2\\
    HD 164507 & 0.380 & 1.104 & 0.785 & 2\\
    HD 3651 & 0.501 & 1.030 & 0.623 & 2\\
    HD 4747 & 0.426 & 1.054 & 0.803 & 1\\
    HD 33632 & 0.398 & 1.079 & 0.803 & 2\\
    HD 130948 & 0.490 & 0.984 & 1.131 & 1\\
    HR 7672 & 0.549 & 1.054 & 0.767 & 1\\
\enddata
\tablerefs{1. \citet{Brewer2016} 2. \citet{Rice2020}}
\end{deluxetable}

\section{Review of Observational Brown Dwarf Spectral Signatures}\label{sec:spectra}
In this section, we review the spectral absorption features for brown dwarfs that have served as observational evidence for the predicted thermochemistry in these atmospheres. These features establish the available data that can be translated into fundamental parameter calculations such as C/O ratio and metallicity in retrieval studies for these objects.

\subsection{Major Absorbing Elements in Ultracool Dwarf Atmospheres}\label{sec: major absorbers}
As established in the foundational works of \citet{Tsuji1964, Lunine1986, Burrows1997, Allard1997, Marley1999a, burrowssharp1999, Lodders1999, Lodders2002, LodFeg2002, Geballe2002, LodFeg2006, Kirkpatrick2005, lodders2010_cloudchem} and \citet{Visscher2010}, the atmospheres of UCDs, and their subsequent spectra, are dominated by C, N, O, Ti, V, Fe, Cr and neutral alkali element chemistry. At temperatures starting near the M/L spectral transition and cooler ($\leq$ 2200 K), we see atomic and neutral atom absorption shift toward broadband molecular absorption features due to H$_2$O, CO, CO$_2$, CH$_4$, NH$_3$, FeH, TiO, VO, CrH, H$_2$S, HCN throughout the optical and infrared \citep[see:][]{Burgasser2002, Geballe2002, Marley2002, Kirkpatrick2005, Cushing2006, Faherty2014, Helling2014}. Retrieval models attempt to constrain the abundances of these absorbers across spectral type -- for L dwarfs mainly H$_2$O, CO, CO$_2$, CH$_4$, FeH, VO, TiO, CrH, Na, K; for T dwarfs H$_2$O, CO, CH$_4$, CO$_2$, NH$_3$, Na, K; for Y dwarfs H$_2$O, CO, CO$_2$, CH$_4$, NH$_3$, PH$_3$. To date, several retrieval studies have been able to constrain abundances of major absorbers (i.e. H$_2$O, CO, CH$_4$, NH$_3$, Na, K) in both L, T and early Y dwarfs \citep[ex:][]{Line2017, Zalesky2019, Zalesky2019b, Burningham2021, Zalesky2022, Calamari2022} with some studies on L dwarfs having shown abundance constraints on minor metal hydrides and oxides (FeH, VO, TiO) as well \citep{Burningham2017, Burningham2021, Vos2023}.

%Some retrieval studies on L dwarfs have succeeded in constraining abundances of both the dominant carbon- and oxygen- bearing species (H$_2$O, CO, CO$_2$, CH$_4$) as well as the minor metal hydrides and oxides (FeH, VO, TiO) \citep[e.g.][]{Burningham2017, Burningham2021, Vos2023}. 
%However, most retrieval models have only been able to constrain major absorbing species (i.e. due to the relative strength of their absorption features throughout the near- and mid-infrared \citep[ex:][]{Line2017, Zalesky2019, Zalesky2022, Calamari2022}.

In this subsection, we provide a brief overview on the dominant thermochemistry in the L, T and Y temperature regimes marked by the most abundant volatile elements (H, C, N, O), as these resulting species drive the C/O ratio and metallicity solutions in current retrieval modelling. We focus our discussion on thermochemical equilibrium assumptions for well-mixed, convective atmospheres as a necessary simplification in our modelling. For a more detailed discussion, see \citet{LodFeg2002}.

\textit{Carbon and Oxygen.} The most identifiable features in UCD spectra often result from carbon and oxygen chemistry in the form of H$_2$O, CH$_4$, CO and CO$_2$. In warmer, less dense atmospheres (i.e., L dwarf atmospheres) we expect to see more carbon and oxygen in CO and CO$_2$ whereas cooler, denser atmospheres (i.e., T or Y dwarf atmospheres) would show more CH$_4$ and H$_2$O.  We can consider CO$_2$ to play a lesser role under thermochemical equilibrium assumptions as it is expected to be observable at much lower pressures ($\log$ P (bar) $<$ -8) than the photospheric pressures we probe (1 $<$ $\log$ P (bar) $<$ 10). For the remaining three major absorbers, they are governed by the net thermochemical reaction:
\begin{equation}
    \rm CO + 3H_2 \rightleftharpoons \rm CH_4 + H_2O
\end{equation}
It is important to note that even in a CO-dominated atmosphere, the abundances of CH$_4$ and H$_2$O do not drop to zero and vice versa. \citet{LodFeg2002} also discuss the implication that overall metallicity ([Fe/H]) has on the CH$_4$ = CO boundary (i.e., the threshold in $P-T$ space between a CO- or CH$_4$-dominated atmosphere). As metallicity decreases, the CH$_4$ = CO boundary shifts to higher temperatures, whereas as metallicity increases the CH$_4$ = CO boundary shifts to lower temperatures. Additionally, while CO$_2$ is not considered a major C-bearing gas in these atmospheres, it is moderately abundant (-12 $< \log X_{\rm CO_2} <$ -6) in a CO-dominated atmosphere. Observational evidence for this species exists in hotter L dwarf atmospheres but subsequent retrieval modelling attempts failed to constrain its relatively low abundance \citep[e.g.][]{Line2015, Gonzales2020} and, as such, is not considered a major contributor in determining C/O ratios for brown dwarfs.

While we focus our analysis in \autoref{sec: predictions} on thermochemical equilibrium assumptions, we do have observational evidence of chemical disequilibrium for the CO to CH$_4$ conversion \citep[i.e.][]{Noll1997, Oppenheimer1998, Miles2020} in brown dwarf atmospheres. This results in higher observed abundances of CO than predicted by thermochemical equilibrium due to rapid vertical mixing from the deeper, hotter atmosphere at rates faster than the chemical timescale conversion to CH$_4$ \citep{Prinn1977, Fegley1996, LodFeg2002, Visscher2006, Visscher2011}. While this is a real, observed phenonmenon, we focus here on thermochemical equilibria for well-mixed atmospheres.

For the CO to H$_2$O conversion, even in the regime where CO is the major C-bearing gas, half of all oxygen can still be in H$_2$O as oxygen is nearly twice as abundant as carbon. It is important to note, as discussed for a solar composition gas in \citet{LodFeg2002}, that the distribution of oxygen atoms between H$_2$O and CO in a UCD atmosphere is going to be affected by the production of oxygen-rich clouds. We discuss theorectical implications of this further in \autoref{sec: theory conds}.

Under equilibrium conditions for pressure and temperature expected in UCD atmospheres (-4 $< \log$ P (bar) $<$ 3; 250 $<$ $T$ (K) $<$ 2500), H$_2$O, CO, CH$_4$ and CO$_2$ will be the most abundant carbon- and oxygen-bearing species. Several other species exist within either CH$_4$- or CO-dominated atmospheres and may play key roles in H-C-N-O reaction kinetics (i.e. CH$_3$, C$_2$H$_6$, CH$_2$O, CH$_3$OH) but their relative abundances and strength of their absorption lines are too weak to consider and have never been recovered in a spectral analysis.

\textit{Nitrogen.} The other major contributor shaping UCD spectra comes from the distribution of nitrogen in NH$_3$. Similar to the carbon and oxygen chemistry above, nitrogen chemistry in these atmospheres is governed by the net thermochemical reaction:
\begin{equation}
    \rm 0.5N_2 + 1.5H_2 \rightleftharpoons \rm NH_3
\end{equation}
In cooler, denser atmospheres, NH$_3$ gas dominates while N$_2$ gas dominates in warmer, less dense atmospheres. Similar to the CH$_4$=CO boundary, the overall metallicity impacts the boundary temperature. As metallicity decreases, the NH$_3$ = N$_2$ boundary shifts to higher temperatures whereas as metallicity increases the NH$_3$ = N$_2$ boundary shifts to lower temperatures. As discussed for the carbon and oxygen chemistry, an N$_2$ dominated atmosphere still has a nonzero abundance of NH$_3$. It is relevant to note here that any object showing NH$_3$ absorption in their spectrum is expected to have CH$_4$ as their major carbon-bearing species \citep{Burrows2003b, Canty2015, Line2017}.

While condensation of N-bearing species into NH$_3$ and/or NH$_4$SH is possible in the coolest atmospheres \citep[e.g., Jupiter and Saturn;][]{Lewis1969, Carlson1987, loddersfegley2006}, these types of condensate clouds are not expected in the warmer atmospheres of L and T dwarfs and do not play a role in subsequent modelling.

As with the carbon chemistry, other minor N-bearing condensates are predicted to exist within either N$_2$- or NH$_3$-dominated atmospheres (i.e., CH$_3$NH$_2$, HCN). As N$_2$ doesn't have absorption features in the near-infrared, its abundance cannot be constrained by observational data and NH$_3$ then remains the only major N-bearing species able to be constrained through retrieval modelling. As a result, NH$_3$ is the only N-bearing species to contribute to metallicity calculations. This hinders our ability to quantify the total nitrogen budget in a given atmosphere.

\subsection{Major Refractory Condensates}\label{sec: theory conds}
Beyond gaseous molecular absorption bands that shape brown dwarf spectra, we have theoretical and observational evidence of condensate absorption as mentioned in previous sections (see \autoref{sec:l dwarfs}, \autoref{sec: major absorbers}). Thermochemically-derived condensation curves of many refractory mineral condensates overlap with the pressure and temperature profiles of UCDs, including CaTiO$_3$, Al$_2$O$_3$, Mg$_2$SiO$_4$, MgSiO$_3$, SiO$_2$, Fe metal, Na$_2$S, Li$_2$S, LiF, KCl, and ZnS \citep{Marley1999a, Chabrier2000, Ackerman2001a, Lodders1999, Marley2002, Lodders2002, LodFeg2002, LodFeg2006, Visscher2006, Visscher2010, Morley2012, wakeford2017}.

While the salt and sulfide clouds are expected in the observable photospheres of cooler T dwarfs, we have yet to find strong spectral evidence of these clouds using retrieval analysis. This could be due to a variety of reasons, including but not limited to clouds sinking below the photosphere in T dwarfs \citep[e.g.,][]{Marley1997, Marley2013, Line2015, Zalesky2022, Calamari2022} or weak or nonexistent spectral absorption features in the near infrared despite observed variance in infrared T dwarf colors \citep{Morley2012}. %and/or low sulfur fugacity disfavoring condensate formation \citet{}.
In warmer objects, or in deeper, hotter layers in T dwarfs, we find the mineral oxide and atomic iron condensates. These condensates will form the most substantial cloud layers in UCD atmospheres due largely to the high relative abundance of magnesium, iron and silicon.

Due to the relatively high condensation temperature of Fe metal, nearly all of this elemental reservoir is condensing into a cloud layer below the photosphere in L dwarfs. This limits Fe abundance above this cloud layer and prohibits it from being a major gaseous absorber (as FeH, FeOH, Fe(OH)$_2$, FeS, etc.;~\citealt{Allard1997, Burgasser2002, burrowssharp1999, Visscher2005, LodFeg2006, Visscher2010}) or a major oxygen sink (as condensed FeSiO$_3$ or Fe$_2$SiO$_4$, iron endmembers of the pyroxene and olivine mineral groups, respectively; \citealt{Visscher2010}). Moreover, the hydrogen-rich atmospheres of UCDs are expected to be too reducing (i.e., the oxygen fugacity is too low) to allow for any appreciable formation of Fe oxides or Fe silicates under equilibrium conditions.

Prior to the condensation points of forsterite (Mg$_2$SiO$_4$) and enstatite (MgSiO$_3$), the magnesium endmembers of the pyroxene and olivine mineral groups, corrundum (Al$_2$O$_3$) and perovskite (CaTiO$_3$) will condense. This will lower the available oxygen inventory above these clouds but as the abundances of calcium, aluminum and titanium are roughly two to three orders of magnitude less abundant than oxygen, the impact on the oxygen inventory is minimal. This is true also for the calcium silicate species anorthite (CaAl$_2$Si$_2$O$_8$) and diopside (CaMgSi$_2$O$_6$) that are condensing in $P-T$ space nearer to forsterite and enstatite but are still limited by the total calcium and aluminum abundance, minimally contributing to the depletion of atmospheric oxygen. So, we are effectively left with condensates that are significant sources of oxygen sinks in UCD atmospheres: forsterite, enstatite and quartz. At present, we have mid-infrared observational evidence of silicate condensate (Mg$_2$SiO$_4$, MgSiO$_3$ and SiO$_2$) absorption \citep{Cushing2006, Burningham2021, Suarez2022, Grant2023}, however, deciphering the exact species responsible remains a challenge.

In the following sections, we focus on the totality of these oxygen-rich condensate species (Mg$_2$SiO$_4$, MgSiO$_3$ and SiO$_2$, Al$_2$O$_3$, CaTiO$_3$, CaAl$_2$Si$_2$O$_8$, CaMgSi$_2$O$_6$) not only because their existence is most feasibly modeled \citep[e.g.][]{Burningham2021, Vos2023} but because they directly tie in to the determination of oxygen abundance in UCD atmospheres. As oxygen abundance can be a potential formation and evolution tracer \citep[see][]{Oberg2011, Madhusudhan2012}, it is essential to understand the impact condensate formation has on oxygen sequestration and how we can most accurately account for this in our modelling.

\section{Theoretical Framework for Thermochemical Analysis in Brown Dwarfs}\label{sec: framework}
One major assumption in the framework of current UCD atmospheric modelling is the use of solar abundance ratios as the standard for understanding the chemistry of these atmospheres beyond spectral line absorbers. These solar abundance ratios are used not only to disentangle the effects of condensate formation but also as population calibrators to help us ground the fundamental parameters of nearby brown dwarfs. However, we've already seen from \citet{Brewer2016}, as illustrated in \autoref{fig:sample}, that the solar C/O ratio lies at the higher end of the population of F, G and K type stars in the local solar neighborhood. This might suggest that using the solar C/O ratio as a chemical marker for the local brown dwarf population is an overestimation. As shown in \citet{Calamari2022}, $>$ 80$\%$ of the current population of retrieved brown dwarfs has a C/O ratio greater than solar ($\sim$ 0.55 from \citet{Lodders2021}) with $>$ 40$\%$ being greater than 0.8, a category of carbon-rich stars thought to be less than $\sim$ 1$\%$ of the local solar neighborhood \citep{Brewer2016b}.

While abundance ratios are a useful trend guide, if we want to understand the chemical makeup of UCD atmospheres it is more informative to examine actual elemental abundances (the occurrence of a given element relative to all other elements). We again point back to \citet{Calamari2022}, where oxygen, specifically, appeared to be depleted in Gl 229B as compared to Gl 229A. While still an open question, this pinpoints what types of atmospheric, or even formation, dynamics could be causing such an outcome. If we want to explore this open question regarding oxygen in brown dwarfs, we turn not only to abundance ratios but individual element abundances. We know that the chemical makeup of the solar neighborhood does vary -- in addition to using broad metrics like C/O ratio and metallicity ([Fe/H]), we want to know how specific element abundance may change and how that can impact subsequent cloud formation. Specifically, we look at species that act as potential oxygen sinks in UCD atmospheres (see \autoref{sec: theory conds}).

%We show abundance ratio trends in \autoref{fig:sample} but in \autoref{fig:individual abundances} we show the variance in fractional (mass-weighted) abundance of relevant refractory condensate forming atomic species.

\begin{figure*}[ht!]
\centering
\includegraphics[width=\textwidth]{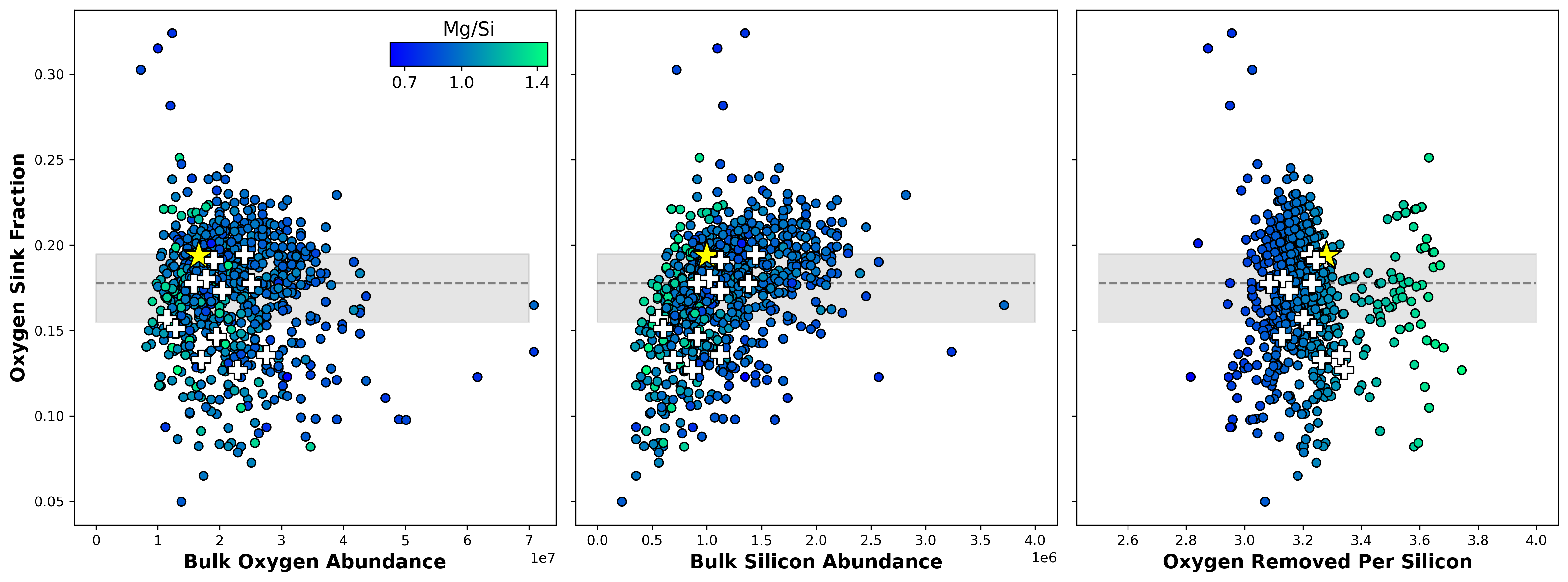}
\caption{Oxygen sink fraction (O$_{\rm sink}$) as function of bulk oxygen abundance (left), bulk silicon abundance (middle) and oxygen removed per silicon (right). Circles represent the solar neighborhood sample from \citet{Brewer2016} colored by their Mg/Si ratio. White crosses indicate the subset of compositional benchmarks and the yellow star represents the Sun. The grey dashed line shows the median oxygen sink fraction for the total solar neighborhood population with the shaded region bounding the first and third quartiles.
\label{fig:o sink frac}}
\end{figure*}

\subsection{Stellar Abundances as a Tool for Understanding Companion Atmospheres}\label{sec: abundances}
By focusing on well-studied primary stars, we can examine the total chemical makeup of a given system by using observationally measured abundances of certain elements. Assuming co-evality, we can use the stellar abundances of both volatile and refractory elements to examine how oxygen is theoretically being sequestered into refractory condensates in a companion UCD atmosphere. This methodology works uniquely for compositional benchmark systems by utilizing known host star element abundances and assuming a similar chemical makeup for its companion. This assumption is strongly supported by the observation in our own solar system that element abundances in the solar photosphere (with the exception of H and He) closely match the element abundances directly measured in the most primitive chondritic meteorites, widely thought to be the building blocks of the planets in our solar system \citep{Lodders2021}. While this introduces a new assumption, we use this method in order to revisit the calculation done in \cite{burrowssharp1999} where they employed solar abundances to predict that approximately 3.28 oxygen per silicon atom would end up in clouds (accounting for $\sim$ 14$\%$ of bulk oxygen). We address implications and evaluations of our assumption in \autoref{sec: future work}.

In order to constrain the amount of total oxygen that would sink into clouds for each benchmark system, we use the published abundances from \citet{Brewer2016, Rice2020} to stoichiometrically calculate how much oxygen will bond with the refractory elements Mg, Si, Ca, Al, Ti, and V. Under thermodynamic equilibrium conditions, in cooler atmospheres, O-bearing condensates form at the transition from the deep interior to cooler pressure layers near the photosphere. In these calculations, we assume that the total bulk abundance of each of these refractory elements is bonding with oxygen, forming MgO, SiO$_2$, CaO, Al$_2$O$_3$, TiO$_2$, VO. Subsequent refractory condensates (see \autoref{sec: theory conds}) can be constructed from combinations of these metal oxides, making the quantitative oxygen sink path-independent -- i.e. the resulting phase composition(s) of the clouds are irrelevant. For example, enstatite and forsterite can be made by combining metal oxide building blocks via net thermochemical reactions:
\begin{equation}
    \rm  MgO + SiO_2 \rightleftharpoons MgSiO_3
\end{equation}
\begin{equation}
   \rm  2 MgO + SiO_2 \rightleftharpoons Mg_2SiO_4
\end{equation}
wherein the total amount of oxygen that may be sequestered into Mg-silicates is determined by the available abundances of Mg and Si, and not upon the relative proportions of particular silicate phases. For a more in-depth discussion of how these various gaseous oxides form mineral condensates, see \citet{Lodders1999, Ackerman2001a, Allard2001, Lodders2002, lodders2010_cloudchem, Visscher2010, wakeford2017}. 

If we consider these types of chemical pathways for both major (Mg, Si) and minor (Ca, Al, Ti, V) refractory elements, we can determine a maximum oxygen sink fraction (the percent fraction of oxygen in condensate clouds) based upon oxidation stoichiometry \cite[e.g., see][]{Visscher2005}:
\begin{equation}
    \Sigma \rm O_{cloud} = 2\Sigma Si + \Sigma Mg + \Sigma Ca + 1.5\Sigma Al + 2\Sigma Ti + \Sigma V
\end{equation}
\begin{equation} \label{eq: osink_fraction}
    \rm O_{sink} = \frac{\Sigma O_{cloud}}{\Sigma O}
\end{equation}
where $\Sigma \rm O_{cloud}$ is the total amount of oxygen taken into metal oxides (and thus condensate clouds), $\Sigma \rm O$ is the total amount of oxygen, and $\rm O_{sink}$ is the fraction of total oxygen in clouds in a given atmosphere. We note that though titanium and vanadium are strong gaseous absorbers, they are typically present in trace ($\sim$ 1$\%$) amounts in the most abundant mineral condensates. However, we include their abundances for completeness in our calculation.

The advantage of this stoichiometric approach is that the total oxygen removal is limited only by the abundances of the major refractory elements and does not require prior knowledge of the distribution of elements into specific condensate phases. While our determination of $\rm O_{\rm sink}$ includes minor refractory elemental abundances (Ca, Al, Ti, V), Mg and Si are responsible for $>$ 90$\%$ of the oxygen removal into such clouds. Moreover, the abundances of minor metals tend to increase with increasing abundances of Mg and Si. Using elemental abundances from the \citet{Brewer2016} solar neighborhood sample, we can thus make a first order approximation for $\rm O_{\rm sink}$ in companion objects using the Mg and Si abundances as a proxy for all metal oxides:
\begin{equation}\label{eq: osink_sio_mgo}
    \rm O_{sink} \approx 2.024(\Sigma Si/\Sigma O) + 1.167(\Sigma Mg/\Sigma O)
\end{equation}
From this relation, the number of oxygen atoms removed per silicon atom (cf.~Fig. \ref{fig:o sink frac} and \citealt{burrowssharp1999}) can also be estimated:
\begin{equation}\label{eq: ocloud/si_removal}
    \rm O_{cloud}/\Sigma Si \approx 2.024 + 1.167(\Sigma Mg/\Sigma S i)
\end{equation}
where $\Sigma \rm Mg/\Sigma Si$ describes the bulk Mg/Si abundance ratio in the system. The significance of this ratio for silicate phase composition will be explored in \autoref{sec: channon methods}.

We can also consider the impact of the oxygen sink on the observable C/O ratio, such that the removal of oxygen into condensed phases will cause the C/O ratio to become greater above the condensate cloud layers relative to the “below-cloud" (i.e. bulk atmosphere) C/O ratio. This “above-cloud", or observed, C/O ratio can be expressed by:
\begin{equation} \label{eq: co_ratio_general}
    \left(\rm C/O\right)_{\rm obs} = \left( \rm C/O\right)_{\rm bulk} \times \frac{1}{1 - \rm O_{\rm sink}}
\end{equation}
where $\rm (C/O)_{\rm obs}$ is the observed C/O ratio in the upper atmosphere, $\rm (C/O)_{\rm bulk}$ is the bulk atmospheric C/O ratio, and $\rm O_{\rm sink}$ is the percent fraction of oxygen sequestered in clouds, calculated from equation (\ref{eq: osink_fraction}).

By substitution of $\rm O_{\rm sink}$ (equation \ref{eq: osink_sio_mgo}) into equation (\ref{eq: co_ratio_general}), the $\rm (C/O)_{\rm obs}$ ratio can be estimated from the bulk elemental abundances for C, O, Mg and Si:
\begin{equation} \label{eq: co_ratio_specific}
    \left(\rm C/O\right)_{\rm obs}  \approx \frac{\rm \Sigma C}{\Sigma \rm O - 2.024\Sigma Si - 1.167 \Sigma Mg}
\end{equation}

Moreover, the observed abundance trends for $\Sigma$Mg, $\Sigma$Si, $\Sigma$O, and the C/O ratio in the \citet{Brewer2016} solar neighborhood sample were used to derive a more general relationship for finding $\rm (C/O)_{\rm obs}$. From this sample, the abundances of Si and Mg can be roughly approximated by $(\Sigma{\rm Si}/\Sigma {\rm O})\!\sim0.1159(\Sigma {\rm C}/\Sigma {\rm O})$ and $(\Sigma{\rm Mg}/\Sigma {\rm O})\!\sim\!0.1165(\Sigma {\rm C}/\Sigma {\rm O})$. Substitution into the above expression thus yields
\begin{equation} \label{eq: co_ratio_bulk}
     \left(\rm C/O\right)_{\rm obs} \approx \frac{(\rm C/ O)_{bulk}}{1-0.371(\rm C/ O)_{bulk}},
\end{equation}
which allows for estimates of the $\rm (C/O)_{\rm obs}$ ratio using only the bulk $\rm (C/O)_{\rm bulk}$ ratio. This expression can likewise be used to estimate bulk C/O, based upon \textit{observed} values of the C/O ratio by accounting for the sequestration of oxygen into major condensate phases.

\subsection{Classifying Major Condensates}\label{sec: channon methods}
In addition to determining the total percentage of oxygen in condensate clouds, we can also use the element abundances and abundance ratios of refractory elements to predict the type of oxygen-bearing clouds we might expect to see and model in a given atmosphere. Similar to the path-independent, stoichiometric calculations above, we can evaluate the silicate condensate sequence by a series of stoichiometric and mass balance calculations.

Due to the high relative abundance of magnesium and silicon (as compared to calcium, aluminum, titanium or vanadium), the dominant oxygen-bearing condensates in UCD atmospheres are expected to be the well-known “silicates": enstatite (MgSiO$_3$), forsterite (Mg$_2$SiO$_4$) and quartz (SiO$_2$). The inventories of these condensates will be affected (1) by the bulk abundances of their constituent elements (Mg, Si) and also (2) by the abundances of the minor refractory elements (Ca, Al, Ti, V) as these elements will condense into even more refractory oxygen-bearing clouds deeper in the atmosphere and can thus affect the available, “above-cloud" inventory of Mg or Si.

A solar-composition gas will condense magnesium and silicon mostly into enstatite, forsterite (and possibly quartz, see below) with $<$ 20$\%$ of Mg or Si into minor refractory condensates \citep[cf.][]{Lodders2002,Visscher2010}. However, to illustrate the silicate condensation sequence, we first consider a solar-composition gas where forsterite and enstatite are the \textit{only} Mg- and Si-bearing condensates. We can thus determine the inventory of forsterite by the following:
\begin{align}
   \Sigma \rm Mg & = 2A_{\rm  Mg_2SiO_4} + A_{\rm  MgSiO_3} \\
    \Sigma \rm Si & =  A_{\rm Mg_2SiO_4} + A_{\rm MgSiO_3} \\
   A_{\rm Mg_2SiO_4} & = \rm \Sigma Mg - \Sigma Si \label{eq: fo_threshold}
\end{align}
where $A_{\rm X}$ is the abundance of a given condensate species and $\Sigma N$ is the abundance of a given element.

From this example, we can see in a gas where Mg/Si $>$ 1, forsterite effectively serves as a sink for “excess" magnesium. If Mg/Si = 1, no forsterite condenses and the only oxygen-rich condensate is enstatite. If Mg/Si $<$ 1, the above expression gives a nonphysical result (i.e., a mass balance violation) for forsterite. In this case, we have a system characterized by “excess" Si which is able to condense into Mg-free species, namely quartz (SiO$_2$).

However, as previously stated, other Mg- and Si- bearing condensates exist and can impact the available inventories of magnesium and silicon. The most abundant of the minor refractory species are diopside (CaMgSi$_2$O$_6$) and anorthite (CaAl$_2$Si$_2$O$_8$) \citep{Allard2001, Lodders2002, Visscher2010}. We can take the first order approximation example from above and extend it to include these refractory species by assuming that anorthite is the main Al-bearing condensate and diopside is the main Ca-bearing condensate (in the very rare case where 2$\Sigma$Al $>$ $\Sigma$Ca, anorthite becomes Ca-limited and the excess Al can form spinel, MgAl$_2$O$_4$). Assuming complete condensation of Ca and Al, the available inventories of Mg and Si at altitudes where we might consider forsterite, enstatite, or quartz as possible condensates are given by:
\begin{align}
    N_{\rm Mg} &  \approx \Sigma {\rm Mg} - A_{\rm CaMgSi_2O_6}\\
    N_{\rm Si} & \approx \Sigma {\rm Si} - 2A_{\rm CaMgSi_2O_6}- 2A_{\rm CaAl_2Si_2O_8}
\end{align}
The forsterite regime threshold (cf.~Eq.~\ref{eq: fo_threshold}) is then
\begin{equation}\label{eq: fo_threshold_2nd_approx}
     A_{\rm Mg_2SiO_4} \approx \rm \Sigma Mg + \Sigma Ca + 0.5 \Sigma Al - \Sigma Si,
\end{equation}
where a nonphysical value ($A_{\rm Mg_2SiO_4}<0$) again indicates an excess of Si and the formation of SiO$_2$. Using element abundances from the \citet{Brewer2016} solar neighborhood sample, this second-order stoichiometric approximation suggests a condensation regime highly sensitive to the bulk atmospheric Mg/Si ratio ($\Sigma$Mg/$\Sigma$Si). For $\Sigma$Mg/$\Sigma$Si $\gtrsim 0.9$ we anticipate the condensation of enstatite + forsterite, whereas for $\Sigma$Mg/$\Sigma$Si $\lesssim 0.9$ we anticipate enstatite + quartz.

In the following section, we discuss variations in the predicted silicate condensation sequence over a range of observed solar neighborhood stellar abundances via \citet{Brewer2016} and as a function of the Mg/Si ratio.

\section{Sequestered Oxygen in Compositional Benchmark Brown Dwarfs}\label{sec: predictions}
In the following subsections we summarize the results of using stellar elemental abundances in compositional benchmark systems to predict the oxygen sink fraction and the silicate cloud regime. Additionally, we calculate these values for the \citet{Brewer2016} solar neighborhood sample to give a population overview of what we might expect to see for the entire compositional benchmark sample (see \autoref{tab:compositional benchmarks}) as well as brown dwarfs in the local region.

\begin{figure}[ht!]
\centering
\includegraphics[width=0.48\textwidth]{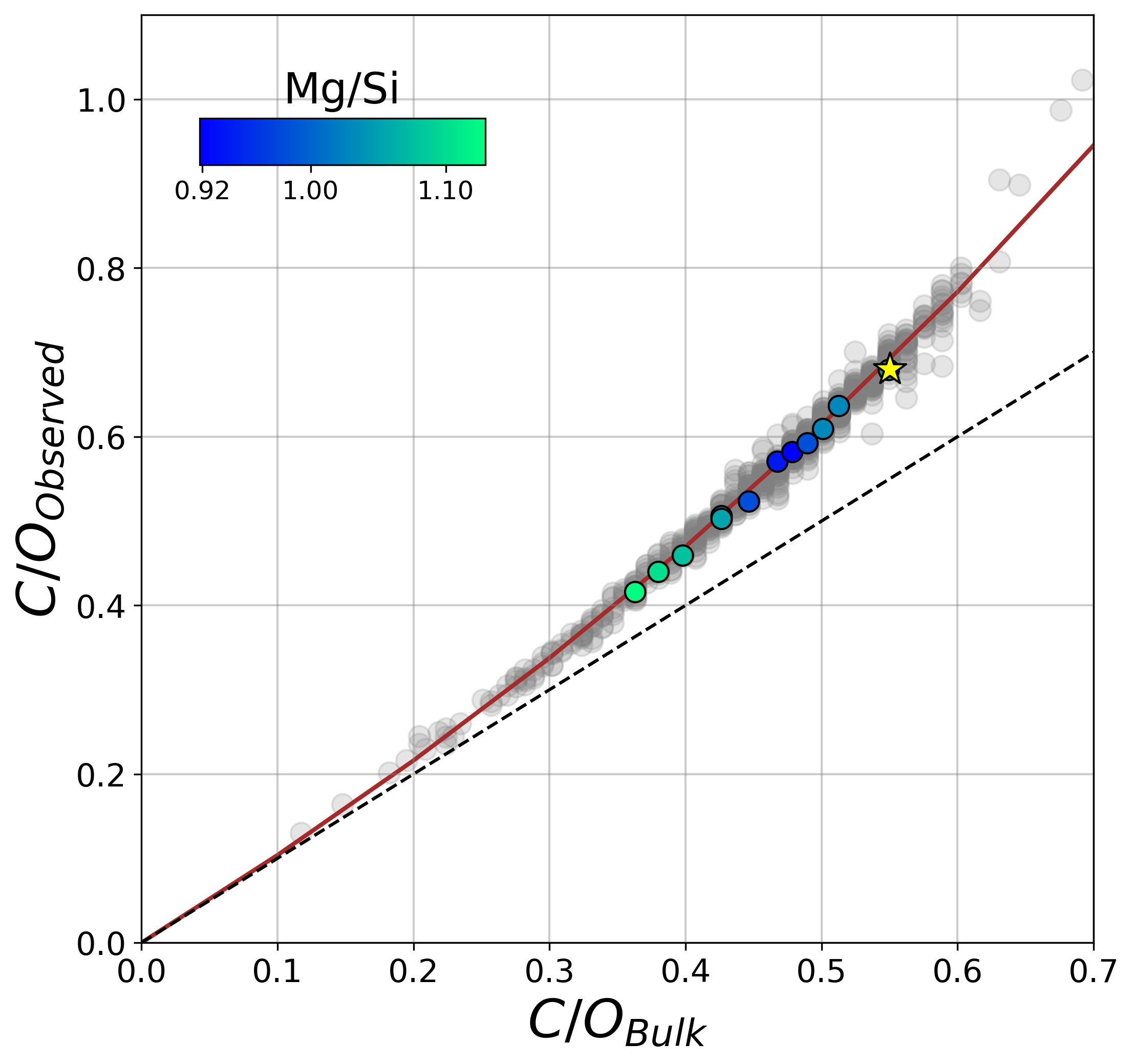}
\caption{The predicted observed C/O ratio in a UCD companion as a function of the system's bulk C/O ratio, using stellar elemental abundances from \citet{Brewer2016} and assuming removal of oxygen via condensation of metal oxides. The dashed line indicates the 1:1 line. The red curve is the estimated $\left(\rm C/O\right)_{\rm obs}$ from equation (\ref{eq: co_ratio_bulk}). The star indicates $\left(\rm C/O\right)_{\rm bulk}$ and the predicted $\left(\rm C/O\right)_{\rm obs}$ ratio using solar elemental abundances from \cite{Lodders2021}. Circles represent the compositional benchmarks subset colored by Mg/Si ratio while grey points show the solar neighborhood sample from \citet{Brewer2016}.\label{fig:c/o ratio}}
\end{figure}

\subsection{Effective Oxygen Removal in Ultracool Dwarf Atmospheres}
Using the mass balance and stoichiometric calculations explained above, we calculate an oxygen sink fraction, or the fraction of bulk oxygen lost to condensates in a UCD atmosphere, for the solar neighborhood as shown in \autoref{fig:o sink frac}. We highlight the individual compositional benchmark systems among this larger sample to illustrate the potential for large variations in chemistry to exist. However, we plot the median of this entire sample such that the chemical distribution of the local solar neighborhood lends itself to an oxygen sink of approximately $17.8^{+1.7}_{-2.3}\%$ (or $\rm O_{\rm sink}\approx 0.178_{-0.023}^{+0.017}$). This is an upper limit estimate by assuming, not unreasonably, that all Mg-, Si-, Ca-, Al-, Ti-, and V-bearing oxides condense out at various points in the UCD atmosphere.

In \autoref{fig:o sink frac}, we show how this oxygen sink varies with bulk oxygen abundance such that in relatively oxygen-rich atmospheres, the refractory elements have a lesser fractional impact where the opposite is true for relatively oxygen-poor environments. In relatively oxygen-poor stars, we find that the bulk abundances of Mg and Si are not necessarily uniformly depleted relative to oxygen thereby sequestering a larger fraction of oxygen in clouds. Additionally, we illustrate that the oxygen sink fraction trends intuitively with bulk silicon abundance such that systems with more silicon will sequester oxygen into clouds at a larger fractional occurrence. Finally, we show the variance in “oxygen removed per silicon atom" -- a parameter defined in \citet{burrowssharp1999} to describe the amount of oxygen in condensates. \citet{burrowssharp1999} estimated an O/Si removal factor of $\sim$ 3.28 using solar abundances from \citet{andersgrevesse1989} while here we illustrate the scatter introduced by using elemental abundances from a larger stellar sample. We find a median O/Si removal factor of $\sim$ $3.19^{+0.05}_{-0.06}$ which intuitively trends with Mg/Si such that systems with higher Mg/Si ratios will have higher O/Si ratios (cf.~equation \ref{eq: ocloud/si_removal}). However, we show that O/Si does not trend with the oxygen sink fraction and is therefore not a useful metric, alone, to estimate the total oxygen removed due to clouds.

The predicted median oxygen sink fraction here is $\sim$ 10 percentage points less than that estimated in previous retrieval modelling work on T-type dwarfs \citep[e.g.][]{Line2015, Calamari2022}, potentially driving the high brown dwarf C/O ratio trend discussed in \citet{Calamari2022} even steeper. If only $\sim$18$\%$ of total oxygen is being lost to clouds, we have to consider what other types of dynamical processes could be occurring in these atmospheres.

Additionally, we show a predictive relation between the above- and below-cloud, or observed and bulk, C/O ratio in \autoref{fig:c/o ratio} as laid out in \autoref{sec: abundances} via \autoref{eq: co_ratio_bulk}. We determine a fit to this data based on the behavior of Mg/O, Si/O and Mg/Si ratios as bulk C/O increases. This generally correlates with a median oxygen sink fraction of $\sim$ 18$\%$ for the solar neighborhood population. However, we do show how observed C/O ratio increases nonlinearly as bulk C/O increases -- a function of the fact that element abundances of O, Mg and Si in metal-poor systems are not necessarily uniformly depleted. As such, as bulk C/O ratio increases, or bulk oxygen abundance decreases (see \autoref{fig:o sink frac}), there is a nonlinear increase in oxygen sink fraction which effects the observed UCD C/O ratio. This fit provides some guidelines for what we might expect given UCD retrieval model outputs.

\subsection{Predicted Silicate Cloud Regime}
\begin{figure}[t!]
\centering
\includegraphics[width=0.49\textwidth]{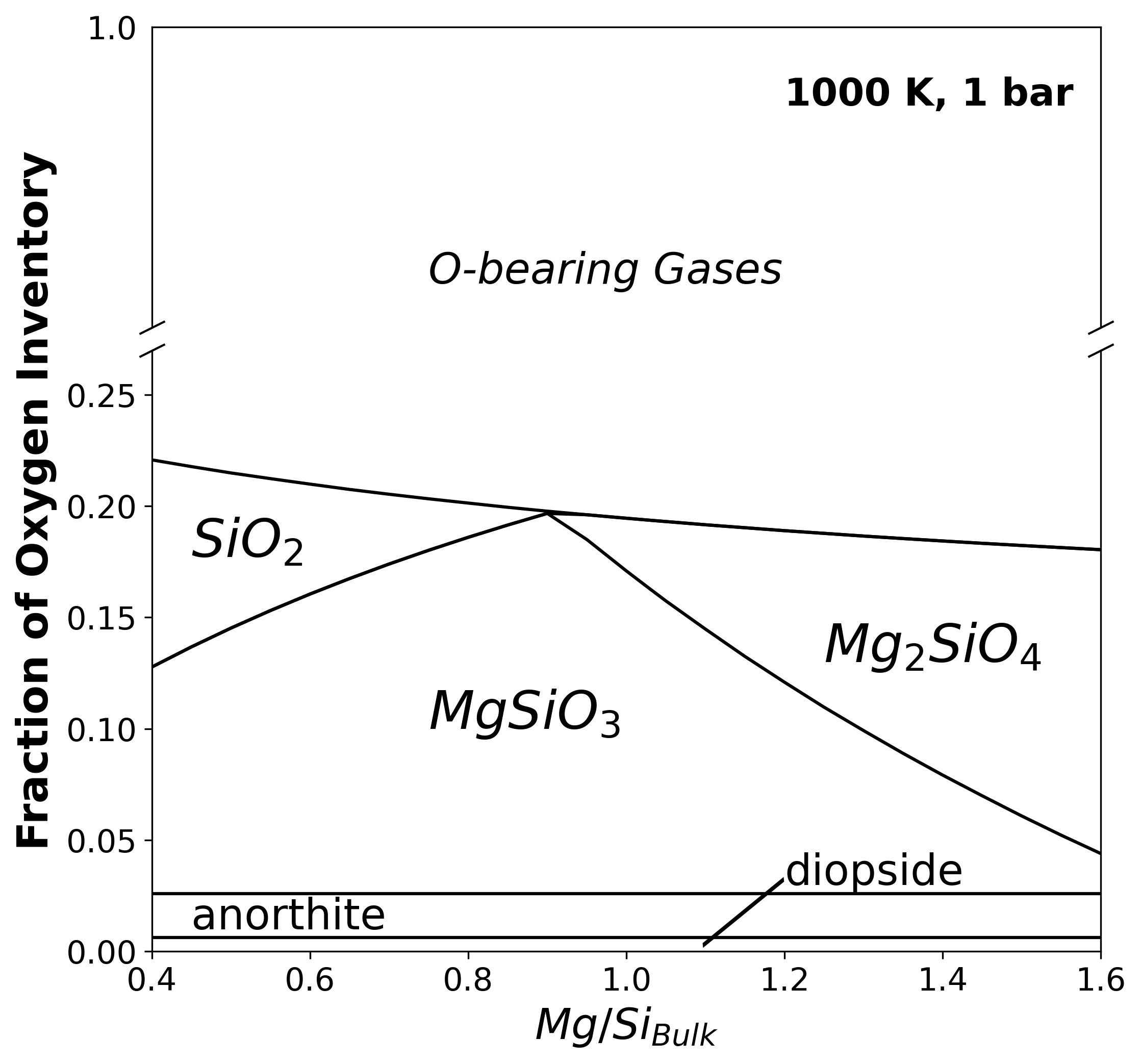}
\caption{Predicted silicate regime and distribution of oxygen into condensates and vapor at 1000 K and 1 bar, based upon thermochemical equilibrium calculations over a range of bulk Mg/Si element abundance ratios in an otherwise solar-composition atmosphere. The fraction of the oxygen inventory removed into condensates corresponds to $\rm O_{\rm sink}$ as described in the text.\label{fig:sil regime}}
\end{figure}

While the range in stellar elemental abundances creates variance in the oxygen sink fraction, it minimally impacts the silicate regime threshold, which is highly sensitive to the Mg/Si ratio. This behavior is demonstrated in \autoref{fig:sil regime}, which shows the equilibrium distribution of O-bearing phases at 1000 K and 1 bar as a function of $\Sigma$Mg/$\Sigma$Si in an otherwise solar-composition gas (solar Mg/Si = 1.03 using abundances from \citealt{Lodders2021}). This $P-T$ point was chosen as it is generally representative of UCD photospheric temperatures and pressures and also lies above Mg-silicate cloud condensation $P-T$ points, thus capturing the upper limit of oxygen sequestration. 

\autoref{fig:sil regime} shows a Mg/Si ratio threshold value of $\sim0.9$, such that UCD companions may be expected to exhibit the following equilibrium silicate condensate regime:
\begin{itemize}
    \item Mg/Si $\lesssim$ 0.9 : Enstatite + Quartz
    \item Mg/Si  $\sim$ 0.9 : Enstatite
    \item Mg/Si $\gtrsim$ 0.9 : Enstatite + Forsterite
\end{itemize}
consistent with the second-order approximation described above (cf.~equation \ref{eq: fo_threshold_2nd_approx}). Moreover, we find that Mg-silicate condensation behavior is relatively independent of varying Ca and Al abundances, etc., suggesting that this threshold may serve as a guide to differentiating silicate cloud regimes in UCD atmospheres. For example, this finding is in agreement with previous retrieval work in \citet{Burningham2021} where the best fit model for the L4.5 dwarf, 2MASSW J2224438-015852, was a layered cloud model consisting of enstatite, quartz and iron with an inferred Mg/Si=0.69.

It is important to keep in mind that the particular sequence and identity of cloud condensate phases in a given UCD atmosphere will be subject to each object's atmospheric properties including element abundance patterns, thermal structure, mixing, condensate re-equilibration, and gravity. Indeed, several works have suggested by closely examining both low and medium resolution optical and near infrared spectra as well as the scatter on color magnitude diagrams, that there is an atmospheric difference in young L dwarfs versus field L dwarfs. This difference can be linked to a low surface gravity in the former (e.g. \citet{Faherty12, Faherty16, Suarez2023}). However, given these possible variations, the stoichiometric approach presented here provides a robust estimate of oxygen removal into refractory condensates and an estimate of atmospheric C/O inventories over a broad range of UCD atmospheres.

\section{Future Applications in Extrasolar World Modelling}\label{sec: future work}
In order to utilize these theoretical predictions, we turn back to retrieval modelling for brown dwarfs, as this is currently the only modelling technique that can explore the unique chemistry and thermodynamics of individualized spectra. These thermochemical predictions will act as guidelines but not constraints in future modelling attempts for compositional benchmarks. In particular, having empirical knowledge about the system will help ground our results in what is already known rather than act as an a priori constraint, potentially biasing results.

In future work, we will specifically return to the compositional benchmark sample outlined in \autoref{tab:sample} and shown in \autoref{fig:o sink frac} and \autoref{fig:c/o ratio}. Of the 12 systems with well characterized host stars, 10 have L-dwarf companions that would be excellent candidates for detailed cloud modelling via retrievals. However, only 4 systems (HD 130948BC, HD 203030B, HR 7672B, HD 4747B) have available near-infrared (NIR) spectral data. Additionally, only 2 of these 4 systems have full NIR spectral coverage (1-2.5 $\mu$m) which has been shown to be the minimum necessary requirement for robust retrieval modelling \citep{Burningham2021}. Beyond NIR spectral coverage alone, \citet{Burningham2021, Calamari2022, Vos2023} demonstrated the need for mid-infrared (MIR) spectral coverage in order to fully model molecular abundances and characterize condensate cloud species. In order to capitalize on the entire compositional benchmark sample outlined in \autoref{tab:compositional benchmarks}, we will continue to require and employ the use of telescopes such as the James Webb Space Telescope (JWST) to provide detailed and full spectral coverage (1-28 $\mu$m) of these objects. Additionally, we will rely on optical telescopes such as HIRES on Keck or PEPSI on the Large Binocular Telescope (LBT) to obtain spectral coverage on the primaries listed in \autoref{tab:compositional benchmarks} to continue characterization of these systems. 

By obtaining full spectral coverage for L-type objects in the compositional benchmarks sample, we can then begin to conduct a suite of models (both cloudy and cloudless) where we might expect a best fit cloud model for each object in this sample to fall into the Mg/Si $>$ 0.9 (enstatite + forsterite) cloud regime. A result counter to that expected would certainly be cause for discussion on the chemical makeup of the UCD companion and its origins.

Additionally, we can apply the oxygen sink correction in future work in the way that has been done in \citet{Line2015, Line2017, Zalesky2022, Calamari2022} -- increasing retrieved oxygen abundance by the percentage lost to clouds. However, this will be a much closer approximation to a true oxygen sink in these atmospheres since we have accounted for system-specific elemental abundances. While our median oxygen sink fraction is a good estimation for oxygen lost to clouds in UCDs that are solitary or have unknown host star chemistry, the compositional benchmark sample outlined in \autoref{tab:sample} has fractions uniquely specific to each system. This subset of host star chemistry reveals variance in oxygen sink fraction from 13 - 19$\%$. Again, we might expect to see trends toward high C/O ratios (or relatively low retrieved oxygen abundances) strengthen as a result of this work. The population of brown dwarfs with a C/O $>$ 0.8 will likely increase. As a result, oxygen loss in brown dwarf atmospheres remains an open question to be explored.

Finally, the work presented here has implications for the atmospheric modeling of gas giant exoplanets whose effective temperatures ($T_{\rm eff}$) cross into the L and T dwarf regime. While exoplanet modeling cannot assume co-evality as we do here given the uncertainty in and influence of the formation process on those worlds, sequestration of oxygen into refractory condensates will certainly impact retrieved molecular abundances in those temperature atmospheres. Despite the added complications of processes such as late-stage accretion, planetary migration and atmospheric differentiation, host star abundance analysis is essential in order to reconstruct formation history. Moreover, the approach described here can provide clues to oxygen sequestration into condensate phases and estimates of bulk composition based on observed abundances of C- and O-bearing species in exoplanet atmospheres.

\section{Conclusions} \label{sec:conclusion}
In this work, we present evidence in favor of using brown dwarfs in comoving systems with F, G or K type stars (“compositional benchmarks") to ground our exploration and understanding of the thermodynamic and chemical processes in brown dwarfs. As F, G and K type stars often have a wealth of data available, this provides us with external empirical information that will help ground our modelling in our attempt to disentangle what the fundamental properties of brown dwarfs, like C/O ratio, are telling us about their atmospheres and formation histories.

Specifically, we have used published elemental abundances for a sample of compositional benchmarks, along with the local solar neighborhood population, from \citet{Brewer2016} to provide us with two empirical constraints: oxygen sink fraction and predicted silicate regime. Through a series of stoichiometric and mass balance calculations, we have determined that, given the bulk elemental abundances from a primary host star, the median oxygen sink in the companion UCD atmosphere is $17.8^{+1.7}_{-2.3}\%$. This update provides context for previous work \cite[e.g.,][]{burrowssharp1999,LodFeg2002, Visscher2005, Visscher2011, Line2015} that have based oxygen sink estimates upon solar elemental abundances. We have also used the elemental abundances of our primary stars to determine the Mg/Si ratio threshold at which the silicate cloud composition transitions from enstatite (MgSiO$_3$) + quartz (SiO$_2$) clouds (Mg/Si $<$ 0.9) to enstatite + forsterite (Mg$_2$SiO$_4$) clouds (Mg/Si $>$ 0.9).

Our global aim in this work is to utilize these chemical predictions in future brown dwarf retrieval modelling studies to help understand the thermochemical dynamics of cloud processes and oxygen sequestration in these atmospheres. By carefully studying brown dwarf atmospheric chemistry, we are one step closer to uncovering the formation and evolution pathways of these objects.

\section*{Acknowledgments}
This work was supported by the National Science Foundation via awards AST-1909776 and AST-1909837 and NASA via award 80NSSC22K0142. This work also benefited from the 2022 Exoplanet Summer Program in the Other Worlds Laboratory (OWL) at the University of California, Santa Cruz, a program funded by the Heising-Simons Foundation. B. Burningham acknowledges support from UK Research and Innovation Science and Technology Facilities Council [ST/X001091/1]. C. Visscher acknowledges support from JWST Cycle 1 AR 2232. The authors thank John M. Brewer for helpful discussions and comments which improved this work.

\bibliography{refs}
\bibliographystyle{aasjournal}

%\appendix
%\restartappendixnumbering
%section{}\label{sec:appendix}

\end{document}